\documentclass[a4paper]{JHEP3}

\usepackage{epsfig,multicol,bbm,amsmath,amssymb}

\newcommand\fverb{\setbox\pippobox=\hbox\bgroup\verb}
\newcommand\fverbdo{\egroup\medskip\noindent%
            \fbox{\unhbox\pippobox}\ }
\newcommand\fverbit{\egroup\item[\fbox{\unhbox\pippobox}]}
\newbox\pippobox

\def\beq{\begin{equation}}
\def\eeq{\end{equation}}
\def\bea{\begin{eqnarray}}
\def\eea{\end{eqnarray}}
\def\beaa{\begin{eqnarray*}}
\def\eeaa{\end{eqnarray*}}

\def\putunder#1#2{\mathrel{
\setbox0=\hbox{#1}\setbox1=\hbox{\scriptsize #2} \dimen0=-0.5\wd0
\advance\dimen0 by -0.5\wd1 \dimen1=0.5\wd0 \advance\dimen1 by
-0.5\wd1
\hbox{\box0\kern\dimen0%
\vbox to 0pt {\hbox{\lower 0.7em \box1}\vss}%
\kern\dimen1} }}

\newcommand{\gsim}{\lower.7ex\hbox{$\;\stackrel{\textstyle>}{\sim}\;$}}
\newcommand{\lsim}{\lower.7ex\hbox{$\;\stackrel{\textstyle<}{\sim}\;$}}
\newcommand{\be}{\begin{equation}}
\newcommand{\e}{\end{equation}}

\title{de Sitter String Vacua from Perturbative K\"ahler Corrections and Consistent D-terms}

\author{Susha L. Parameswaran
        and Alexander Westphal\\
    ISAS-SISSA and INFN, Via Beirut 2-4, I-34014 Trieste, Italy\\
    E-mail: \email{param@sissa.it}, \email{westphal@sissa.it}}

\preprint{SISSA-08/2006/EP\\February 23, 2006}  

\abstract{We present a new way to construct de Sitter vacua in
type IIB flux compactifications, in which moduli stabilization and
D-term uplifting can be combined in a manner consistent with the
supergravity constraints. Here, the closed string fluxes fix the
dilaton and the complex structure moduli while perturbative
quantum corrections to the K\"ahler potential stabilize the volume
K\"ahler modulus in an $AdS_4$-vacuum. Then, the presence of
magnetized $D7$-branes in this setup provides supersymmetric
D-terms in a fully consistent way which uplift the $AdS_4$-vacuum
to a metastable $dS$-minimum.}


\begin{document}


\baselineskip=18pt

\section{Introduction}

The last few years have seen the discovery of a vast
'landscape'~\cite{BoussoP,kklt,sussk,dougl} of stable and
meta-stable 4d vacua of string theory.  This marks remarkable
progress in the formidable task of constructing realistic 4d
string vacua.  In particular, the most pressing issues have been
how to stabilize the geometrical moduli of a compactification, and
at the same time address the tiny, positive cosmological constant
that is inferred from the present-day accelerated expansion of the
universe. Recently, the use of closed string background fluxes in
string compactifications has been studied in this
context~\cite{GKP,CBachas,PolStrom,Michelson,
DasSeRa,TaylVaf,GVW,Vafa,Mayr,GSS,KlebStrass,Curio2,CKLT,HaaLou,BB,
DallAgata,KaScTr,silver,acharya,dlust}. Such flux
compactifications can stabilize the dilaton and the complex
structure moduli in type IIB string theory. Non-perturbative
effects such as the presence of $Dp$-branes~\cite{Verl} and
gaugino condensation were then used by Kachru {\it et al}
(henceforth KKLT)~\cite{kklt} to stabilize the remaining K\"ahler
moduli in such type IIB flux compactifications (for related
earlier work in heterotic M-theory see~\cite{Curio1}).
Simultaneously these vacua allow for SUSY breaking and thus the
appearance of metastable $dS_4$-minima with a small positive
cosmological constant fine-tuned in discrete steps.
KKLT~\cite{kklt} used the SUSY breaking effects of an
$\overline{D3}$-brane to achieve this. Alternatively the effect of
D-terms on $D7$-branes has been considered in this
context~\cite{bkqu}.

Bearing in mind the importance of constructing 4d de Sitter string
vacua in a reliable way, one should note the problems of using
$\overline{D3}$-branes as uplifts for given volume-stabilizing AdS
minima. The SUSY breaking introduced by an $\overline{D3}$-brane
is explicit and the uplifting term it generates in the scalar
potential cannot be cast into the form of a 4d ${\cal N}=1$
supergravity analysis. Thus, the control that we have on possible
corrections in supergravity is lost once we use
$\overline{D3}$-branes for SUSY breaking. Replacing the
$\overline{D3}$-branes by D-terms driven by gauge fluxes on
$D7$-branes~\cite{bkqu} is a way to alleviate this problem because
then the SUSY breaking is only spontaneous. In this case the
requirements of both 4d supergravity and the $U(1)$ gauge
invariance necessary for the appearance of a D-term place
consistency conditions on the implementation of a D-term (noted
in~\cite{bkqu}, and emphasised in~\cite{BDKP,DuVe,VZ}). These
conditions have not yet been met by any concrete stringy
realisation of~\cite{bkqu}, where the proposal was made in the
context of KKLT. A consistent mechanism of stabilizing a modulus
via D-terms and uplifting its minimum to a metastable dS vacuum
has been constructed within the context of 4d supergravity
by~\cite{VZ} without, however, having a viable string embedding -
a more stringy and consistent model can be found in~\cite{DuVe}.

In view of these difficulties it is appealing that recently the
possibility of stabilizing the remaining K\"ahler volume modulus
of type IIB flux compactifications purely by perturbative
corrections to the K\"ahler potential has been
studied~\cite{HG,BHK2}. The leading corrections which the K\"ahler
potential receives are given by an ${\cal
O}(\alpha'^3)$-correction~\cite{bbhl} and string loop
corrections~\cite{BHK1}. The $\alpha'$-corrections have recently
been used to provide a realization of the simplest KKLT $dS$-vacua
without the need for $\overline{D3}$-branes as the source of
uplifting~\cite{Brama,Bobk,West}. Under certain conditions the
interplay of both the $\alpha'$-correction and the loop
corrections leads to a stabilization of the volume modulus by the
perturbative corrections alone~\cite{BHK2}. The corrections to the
K\"ahler potential do not break the shift symmetry of the volume
modulus. Therefore, in the present note, we show that such a
K\"ahler stabilization mechanism allows for a consistent D-term
uplift, by gauging this shift symmetry with world-volume gauge
fluxes on $D7$-brane.  Moreover, from simple scaling arguments one
can conclude that the resulting vacuum does not suffer from any
tachyonic directions.

The paper is organized as follows. Section~\ref{D-term} reviews
the D-term uplifting procedure. Further, it summarizes the known
constraints from 4d ${\cal N}=1$ supergravity on the
implementation of so-called field dependent Fayet-Iliopoulos (FI)
D-terms by gauging a $U(1)$ shift symmetry. In Section~\ref{AdS}
we review the mechanism of stabilizing the volume modulus in an
$AdS_4$-minimum by using the perturbative corrections to the
K\"ahler potential, whose structure is summarized. These results
are then used in Section~\ref{dS} to gauge the $U(1)$ shift
symmetry of the volume modulus by turning on gauge flux on a
single $D7$-brane. Then, for an illustrative - if incomplete -
example, we calculate the full scalar potential resulting from the
F-terms and the D-term and show that by an appropriate tuning of
the fluxes we can obtain a metastable $dS$-vacuum for the volume
modulus with all the other moduli also fixed. Finally, we
summarize our results in the Conclusion.

\section{D-terms uplifts and consistency conditions from 4d
${\cal N}=1$ supergravity}\label{D-term}

The proposal to use a field dependent FI D-term as a source of
uplifting $AdS$- to $dS$-vacua was constructed in~\cite{bkqu}.
Consider a 4d ${\cal N}=1$ compactification of type IIB string
theory on an orientifolded Calabi-Yau 3-fold in the presence of
closed string fluxes. The $G_{(3)}$-flux fixes the dilaton $S$ and
the complex structure moduli $U^I$. Generically, this procedure
leaves the K\"ahler moduli unfixed and in particular the universal
K\"ahler volume modulus $T$. Now, the volume modulus enjoys a
Peccei-Quinn type symmetry: $T \rightarrow T + i\alpha$.  In the
presence of a background 2-form gauge field strength $F_{mn}$,
threading the world-volume of a D7-brane wrapped on a 4-cycle
$\Gamma$ of the compact internal manifold, this symmetry is
gauged.  The corresponding gauge covariant derivative acts on
$b={\rm Im}\,T$ as $D_{\mu}b=\partial_{\mu}b+iqA_{\mu}$, with $q$
the charge.  The necessary coupling, $qA^{\mu}\partial_{\mu}b$,
arises from the $a_{(2)}\wedge F_{(2)}$-coupling contained in the
world volume action of the $D7$-brane, where $b$ and $a_{(2)}$ are
dual fields. Here $a_{(2)}$ denotes the 2-form potential contained
in the closed string 4-form $C_{(4)}$ which has the world volume
coupling $C_{(4)}\wedge F_{(2)}\wedge F_{(2)}$ to the $U(1)$-gauge
field strength $F_{(2)}=dA_{(1)}$ on the $D7$-brane. As long as we
assume just one $D7$-brane its $U(1)$ world volume gauge theory
has no local anomalies.\footnote{In case of a stack of coincident
$D7$-branes there will be in general an anomaly whose cancellation
arises via a generalized Green-Schwarz mechanism in a
model-dependent way. The additional Green-Schwarz coupling,
$q'A^{\mu}\partial_{\mu}b'$, may be small enough compared to the
coupling induced by the world volume gauge fluxes to be
neglected~\cite{Ibanez,Antoniadis:2002cs}.}

As expected, the gauging goes hand in hand with a D-term
potential, and specifically a contribution to the scalar potential
for the volume modulus $T$.  This arises from the world-volume
action of the wrapped D7-brane, as: \beq V_D(T)\sim
T_7\cdot\int_{\Gamma}d^4y\sqrt{g_8}F_{mn}F^{mn} \sim
\frac{q^2}{(T+\bar{T})^3}\;\;,\label{BKQuplift} \eeq where $T_7$
is the $D7$-brane tension. For simplicity we are assuming a single
K\"ahler modulus, and also the absence of matter fields charged
under the $U(1)$ gauge group. This latter assumption may be
justified in a model with a single isolated $D7$-brane: The matter
fields arising from open strings stretching between the $D7$- and
other branes would then become very massive thus driving their
VEVs to zero. In the presence of light charged matter fields, one
must consider whether their dynamics are such as to minimise the
D-term potential at $V_D=0$, or to allow this supersymmetry
breaking contribution.\footnote{See \cite{bkqu} for more
discussion on this important point.}

Since it is a shift symmetry of the chiral superfield $T$ that is
gauged by the $D7$-brane gauge flux, we will summarize now the
requirements of supergravity for the case where the $U(1)$ shift
symmetry of a general modulus $\Phi$ is gauged, and gives rise to
a field dependent FI D-term ~\cite{BDKP,VZ}.  Consider a typical
tree level K\"ahler potential of the form:\footnote{In terms of
the notation in~\cite{wess&bagger}, the action of 4D ${\cal N}=1$
supergravity is specified by the functions $K(T, {\bar T})$,
$W(T)$, $\Gamma(T,{\bar T}, V)$ and $f(T)$.  Some formulations of
the gauged supergravity absorb the function $\Gamma$ into a
modified K\"ahler potential $K'$, which - in the case at hand - is
defined as $K = -3\cdot\ln(T+ {\bar T}) \rightarrow K' = -3
\cdot\ln(T + {\bar T}+cV)$ (see {\it e.g.} \cite{bkqu}).  In the
latter formulation, we have then $D=\frac{\partial K'}{\partial V}
\vline_{V=0}$, which yields the same as below.} \beq
K=-p\cdot\ln(\Phi+\bar{\Phi})\label{KS} \eeq where $p$ is a
constant prefactor, which is $3$ for the volume modulus $T$ and
$1$ for the dilaton $S$. The shift symmetry of $\Phi$ is assumed
to be an isometry of its K\"ahler potential: Under \beq \Phi\to
\Phi+i\alpha\label{shift}\eeq the K\"ahler potential is invariant
up to a K\"ahler transformation. Defining now the function \beq
G=K+\ln|W|^2\label{G}\eeq with $W$ denoting the superpotential,
the scalar potential reads \beq
V=V_F+V_D=e^G(G^{\Phi\bar{\Phi}}G_{\Phi}
\overline{G_{\Phi}}-3)+\frac{1}{2}\,({\rm
Re}\,f_{\Phi})^{-1}D_{\Phi}^2\;\;.\label{Veff}\eeq where
$G_{\Phi}=\partial G/\partial\Phi$, and $f_{\Phi}$ is the gauge
kinetic function.  Here the D-term is given as a solution to the
Killing equation of the isometry eq.~\eqref{shift} by \beq
D_{\Phi}=iX^{\Phi} G_{\Phi}=-\,q\cdot\left(\frac{\partial
K}{\partial\Phi}+\frac{1}{W}\cdot\frac{\partial
W}{\partial\Phi}\right)=q\cdot\left(\frac{p}{\Phi+\bar{\Phi}}-
\frac{W_{\Phi}}{W}\right) \label{Dterm}\eeq where $X^{\Phi}=iq$
denotes the Killing vector of the isometry eq.~\eqref{shift}. The
requirement for this D-term to exist is that the shift symmetry of
$\Phi$ is promoted to a $U(1)$ gauge symmetry of the full
supergravity. Gauging of the shift symmetry requires $G$ to be
invariant under the isometry. This results in the most general
form of the superpotential $W$ consistent with gauge invariance
being \beq W=A\cdot e^{a\Phi}\label{superpot}\;\;,\eeq for some
constants $A$, $a$. If $W$ is independent of $\Phi$ and the gauge
kinetic function has a typical stringy modular dependence
$f_{\Phi}=\Phi$ then the D-term generates a potential $V_D\sim
(\Phi+\bar{\Phi})^{-3}$ of the type of eq.~\eqref{BKQuplift}.

In type IIB string theory compactified to 4d the role of $\Phi$ is
played by the universal K\"ahler volume modulus $T$. However, the
superpotential in type IIB flux compactifications (for the case of
just the one universal K\"ahler modulus $T$, i.e. the volume)
including non-perturbative effects generically takes the form \beq
W=W_{\rm flux}(S,U^I)+A\cdot e^{a T}\label{WIIB}\;\;.\eeq This
form of the superpotential is guaranteed to all orders in
perturbation theory by a non-renormalization
theorem~\cite{Burgnonrenorm}. Unless the non-perturbative
corrections in $T$ are absent, this superpotential breaks the
invariance under an isometry of the type of eq.~\eqref{shift}
which the tree level K\"ahler potential of $T$ has. Thus, if
background fluxes are used to stabilize $S$ and the $U^I$ and
non-perturbative effects are used to stabilize $T$, as in
KKLT~\cite{kklt}, then the shift symmetry that the K\"ahler
potential has cannot be gauged to yield a D-term
uplift~\cite{VZ}.\footnote{See, however, \cite{Carlos}.}

In the absence of non-perturbative effects, the shift symmetry of
$T$ can be promoted to a gauge symmetry, thus providing a D-term
uplift of precisely the form of eq.~\eqref{BKQuplift}. Here we are
assuming, for simplicity, that the $U(1)$ gauging the shift
symmetry lives on a single $D7$-brane wrapped on a 4-cycle
$\Gamma$ with ${\rm Re}\,T={\rm Vol}(\Gamma)$, which implies
$f_T=T$. Indeed, requiring that none of the wrapped D-branes are
stacked guarantees that there is no gaugino condensation in
non-Abelian gauge sectors with gauge coupling $T$.  This
restriction in brane distributions amounts to a choice of flux.
Alternatively one can hope to avoid gaugino condensation by noting
that, in fact, world-volume gauge theories usually have too much
matter to generate superpotentials~\cite{explicitmodels}.
Similarly, Euclidean $D3$-brane instantons - which would also
break the shift symmetry - are not a generic
phenomenon~\cite{WittenInst}. Since both effects are exponentially
suppressed they would anyway break the shift symmetry only on a
much lower scale than the typical scale of perturbative
corrections.\footnote{In any case, gaugino condensation on stacks
of D7's should not pose any problem if we have more than one
K\"ahler modulus. Rather it could just serve to lift the
additional K\"ahler moduli, which correspond to the size of their
wrapped 4-cycles $\Gamma^i$, $T_i$.  This can be seen from the
effective 4D gauge coupling, which descends from the world-volume
coupling as $\int_{\mathbb{R}^4\times\Gamma^i}\sqrt{g_8}\cdot
F_{\mu\nu}^iF^{\mu\nu\;i}
\sim\int_{\mathbb{R}^4}\sqrt{\tilde{g}_4}
\cdot(T_i+\bar{T}_i)F_{\mu\nu}^iF^{\widetilde{\mu\nu}\;i}$. Here a
tilde denotes the use of a rescaled metric without dependence on
the warp factor. Meanwhile, world-volume background fluxes could
still gauge the (thus far unbroken) shift symmetry in the volume
modulus, as seen from $a_{(2)} \wedge F_{(2)}^i \int_{\Gamma^i}
J_{(2)} \wedge F_{(2)}^i$.} Keeping the invariance under shifts
(to allow the D-term) while stabilizing $T$ demands that $T$ has
to be stabilized by corrections depending solely on $T+\bar{T}$.
By holomorphy of the superpotential we are then led to consider
stabilization of $T$ by perturbative corrections to the K\"ahler
potential, which depend only on $T+\bar{T}$.

\section{Perturbative corrections to the K\"ahler potential and
volume stabilization}\label{AdS}

Recently the possibility of stabilizing the volume modulus of type
IIB flux compactifications solely by perturbative corrections to
the K\"ahler potential has received some
attention~\cite{bbhl,HG,BHK2}. This is due to the fact that the
two leading corrections have been derived in type IIB string
theory explicitly (for a few concrete examples, at least).

Firstly, one has in type IIB compactified on an orientifolded
Calabi-Yau threefold an ${\cal O}(\alpha'^3)$ $R^4$-correction to
the 10d type IIB supergravity action~\cite{bbhl,GreenSethi} (see
below for a discussion of other corrections at ${\cal
O}(\alpha'^3)$) \bea
S_{\textrm{IIB}}&=&-\frac{1}{2\kappa_{10}^2}\int
d^{10}x\sqrt{-g_{\textrm s}}\;e^{-2\phi}\left[R_{\textrm
s}+4\left(\partial\phi\right)^2+\left.\alpha^{\prime}\right.^3
\frac{\zeta(3)}{3\cdot
2^{11}}\;J_0\,+\,\ldots\,\right]\;\;.\label{aprim}\eea Here $J_0$
denotes the higher-derivative interaction {\small
\[J_0=\left(t^{M_1N_1\cdots M_4N_4}
t_{M_1^{\prime}N_1^{\prime}\cdots M_4^{\prime}N_4^{\prime}}
\hspace{-0.3ex}+\hspace{-0.5ex}\frac{1}{8}\;\epsilon^{AB
M_1N_1\cdots M_4N_4}\epsilon_{AB M_1^{\prime}N_1^{\prime}\cdots
M_4^{\prime}N_4^{\prime}}\right)\hspace{-1ex}
\left.R^{M_1^{\prime}N_1^{\prime}}\right._{M_1N_1}\cdots
\left.R^{M_4^{\prime}N_4^{\prime}}\right._{M_4N_4}\;,\]} and the
tensor $t$ is defined in~\cite{tensort}.  This generates a
correction to the K\"ahler potential \bea \Delta
K_{\alpha'^3}^{R^4}&=&-2\cdot
\ln\left(1+(2\pi\alpha')^3\frac{\hat{\xi}}{2{\cal
V}}\right)\;,\;\;\;
\hat{\xi}=-\frac{1}{4\sqrt{2}}\;\zeta(3)\cdot\chi\cdot(S+\bar{S})^{3/2}
=:\xi\cdot(S+\bar{S})^{3/2}\nonumber\label{dK}\\
&=&-2\cdot\ln\left(1+(2\pi\alpha')^3
\frac{\hat{\xi}}{2(T+\bar{T})^{3/2}}\right)\nonumber\\
&=&-(2\pi\alpha')^3\frac{\hat{\xi}}{(T+\bar{T})^{3/2}}+{\cal
O}(\alpha'^6)\;\;.\eea Here ${\cal V}=(T+\bar{T})^{3/2}$ denotes
the Calabi-Yau volume and from now on we set $2\pi\alpha'=1$ (of
course one can put the appropriate powers of $2\pi\alpha'$ into
the final results by using dimensional analysis).

Next, there exist string loop corrections to the K\"ahler
potential. Ref.~\cite{HG} studied field theory loop corrections
arising in the 4d ${\cal N}=1$ supergravity after compactification
of type IIB string theory, which by dimensional analysis start
with a correction to the K\"ahler potential $\sim
(T+\bar{T})^{-2}$. The string loop corrections have been
calculated explicitly by~\cite{BHK1} for compactification of type
IIB string theory on the $T^6/\mathbb{Z}_2\times\mathbb{Z}_2$
orientifold with Hodge numbers $(h_{11},h_{21})=(3,51)$, and for
the ${\cal N}=2$ sector contribution in the $T^6/\mathbb{Z}_6'$
orientifold. Here, the induced corrections to the K\"ahler
potential have the form \beq \Delta
K^{(g_s)}=-\beta_1\cdot\frac{\mathcal{E}_2^{\rm
D3}(A,U)}{(S+\bar{S})(T+\bar{T})}-\beta_2
\cdot\frac{\mathcal{E}_2^{\rm
D7}(0,U)}{(T+\bar{T})^2}\;\;,\label{dK4}\eeq where
$\mathcal{E}_2^{\rm D3}(A,U)$ and $\mathcal{E}_2^{\rm D7}(0,U)$
are string loop functions, depending on the complex structure
moduli, collectively denoted with $U$, and D3-brane position open
string moduli, $A$.\footnote{The D7-brane scalars and the twisted
moduli have been neglected.} $\beta_1$ and $\beta_2$ are constants
which read $\beta_2=\beta_1=3/256\pi^6$ on the
$T^6/\mathbb{Z}_2\times\mathbb{Z}_2$ while for the
$T^6/\mathbb{Z}_6'$ they have not been determined~\cite{BHK1}.
Assuming that the dilaton and complex structure are stabilised,
$D_S W = 0$ and $D_U W = 0$, then the large volume expansion of
the scalar potential induced by all the above corrections starts
with~\cite{HG,BHK2} \beq
V=e^{K^{(0)}}\cdot|W|^2\cdot\left(\frac{c_1}{(T+\bar{T})^{3/2}}+
\frac{c_2}{(T+\bar{T})^2}+\ldots\right) \label{VlargeT}\eeq where
the first contribution comes from the $\alpha'$ correction, and
the second from the string loop correction. Here, the constants
$c_1$ and $c_2$ are given by: \beq c_1 =
3/4\cdot\hat{\xi}\;\;,\label{c1} \eeq \beq c_2 = 2\,\beta_2\cdot
\mathcal{E}_2^{\rm D7}(0,U)\;\;.\label{c2}\eeq Note that it is
again the piece $\sim(T+\bar{T})^{-2}$ in $\Delta K^{(g_s)}$ which
is the relevant loop correction.\footnote{Terms in $\Delta
K^{(g_s)}$ scaling as $(T+\bar{T})^{-1}$ cancel out to leading
order in the scalar potential. Here, their contribution at ${\cal
O}((T+\bar{T})^{-2})$ (in the bracket of eq.~\eqref{VlargeT}) is
suppressed by $({\rm Re}\,S)^{-2}$ which allows us to consistently
drop it.\label{d3cont}}

Now we can see that when $c_2>0$ and $c_1<0$ (which corresponds to
$\chi>0$) and $|c_2/c_1|\gg 1$ there is inevitably a
non-supersymmetric $AdS_4$-minimum for the scalar potential of
${\rm Re}\,T$ containing both corrections at large
volume~\cite{HG} (see also~\cite{BHK2}).~\footnote{For an
alternative approach to K\"ahler stabilization without turning to
non-perturbative effects see~\cite{SKhalil}, where, however, the
gauge consistency conditions for D-terms from 4d ${\cal N}=1$
supergravity were not discussed.}

Unfortunately, in the only fully calculated example,
$T^6/\mathbb{Z}_2\times\mathbb{Z}_2$, we have
$\chi=2\cdot(h_{11}-h_{21})<0$, for which there is no minimum. We
may however look to the orientifold
$T^6/\mathbb{Z}_6'$~\cite{BHK1,BHK2} as a promising candidate for
the implementation of our scenario. There, $\chi>0$ and the known
${\cal N}=2$ part of the loop corrections takes the same form as
the $T^6/\mathbb{Z}_2\times\mathbb{Z}_2$ corrections (the
inequivalent $T^6/\mathbb{Z}_2\times\mathbb{Z}_2$-orientifold also
has $\chi
>0$ but there the requirement of exotic $O$-planes~\cite{AntLoop}
may complicate the loop corrections, which are presently unknown).

We therefore review the semi-explicit example of type IIB on the
$T^6/\mathbb{Z}_6'$-orientifold discussed in~\cite{BHK1,BHK2},
neglecting the unknown ${\cal N}=1$ sector contribution for the
moment. This example will serve to illustrate the dynamics,
although numerical results may change in a complete calculation of
this - or any other - model.  This orientifold has besides the
volume modulus $T$ one untwisted complex structure modulus $U$ and
the dilaton $S$. The K\"ahler potential including the corrections
reads (we drop the $D3$-brane contribution - see footnote
\ref{d3cont}) \beq
K=-3\cdot\ln(T+\bar{T})-\ln(U+\bar{U})-\ln(S+\bar{S})
-\frac{\hat{\xi}}{(T+\bar{T})^{3/2}}
-\beta_2\cdot\frac{\mathcal{E}_2^{\rm
D7}(0,U)}{(T+\bar{T})^2}\label{Kfull}\eeq where for ${\rm
Re}\,U\gg 1$ we have $\mathcal{E}_2^{\rm
D7}(0,U)=1/\beta\cdot(U+\bar{U})^2$ with $\beta$ a constant,
assumed positive.  Combined with the flux superpotential \beq
W=\frac{1}{2\pi}\,\int_{\cal M}G_{(3)}\wedge
\Omega\label{Wflux}\eeq this fixes $U$ and $S$ in minima given by
the conditions $D_UW=0$ and $D_SW=0$.\footnote{Note that the
string loop corrections of~\cite{BHK1} have been calculated in the
absence of bulk fluxes. As usual, we assume that the back reaction
of the $G_{(3)}$-flux on the geometry is weak in the large volume
limit, which leaves the corrections unchanged to leading order.
Possible additional corrections from the open string/D-brane
sector have also been neglected in~\cite{bbhl, BHK1}.} We then
arrive at a scalar potential for $T$ given by \beq
V_F(T)=e^{K^{(0)}}|W|^2\cdot\left(\frac{3}{4}\,
\frac{\hat{\xi}}{(T+\bar{T})^{3/2}} +\frac{2\beta_2}{\beta}\,
\frac{(U+\bar{U})^2}{(T+\bar{T})^2}\right)\;\;.\label{VF}\eeq
Since $\chi=48>0$ on the $T^6/\mathbb{Z}_6'$ this leads ($c_1<0$
and $c_2>0$ here) to an $AdS_4$-minimum for ${\rm Re}\,T$ at \beq
{\rm Re}\,T=\left(\frac{80\,\beta_2}{27\,\beta}\right)^2
\cdot\frac{1}{\xi^2}\cdot\frac{({\rm Re}\,U)^4}{({\rm
Re}\,S)^3}\;\;.\label{Tmin}\eeq From this expression it is clear
that tuning the $AdS_4$-minimum for ${\rm Re}\,T$ to moderately
large volume needs a tuning of ${\rm Re}\,U$ to large values
\cite{BHK2}.

Let us now comment on the large volume expansion in presence of
the other higher-dimension operators appearing in the type IIB
effective action at ${\cal O}(\alpha'^3)$. The full supersymmetric
form of the $\alpha'$-correction in 10d is interpreted
in~\cite{bbhl,CQS} as arising at the bosonic level from both the
$R^4$-term and $R^3G^2$- and $R^2(DG)^2$-corrections. The form of
the contribution to the scalar potential induced by eq.~\eqref{dK}
shows a volume scaling \beq \Delta V_{\alpha'^3}^{R^4}\sim{\cal
V}^{-3}\label{volscaling1}\eeq compared with the scaling of the
tree level flux contribution \beq V_{\rm flux}=e^{K^{(0)}}\cdot
K_{cs}^{I\bar{J}}D_IW \overline{D_JW}\sim {\cal
V}^{-2}\label{volscaling2}\;\;.\eeq Here $K_{cs}$ denotes the tree
level K\"ahler potential for the complex structure moduli. All
other competing higher-dimension operators at ${\cal
O}(\alpha'^3)$ (such as $G^4R^2\,,\;\ldots\,,\;G^8$) have a
subdominant volume scaling $\sim{\cal V}^{-s}\,,\;s\geq 11/3$ with
respect to the leading correction discussed above~\cite{CQS}. Note
that this is also subdominant with respect to the string theoretic
1-loop corrections given above since they induce a correction to
the scalar potential scaling as \beq \Delta
V_{g_{\hspace*{-0.1ex}\rm S}}\sim{\cal
V}^{-10/3}\;\;.\label{volscaling3}\eeq Furthermore, within our
assumptions, higher-order $\alpha'$-corrections are subleading to
the above string loop corrections in the region of the scalar
potential around the minimum.

Finally, we should comment on the stability of the minimum found
above with respect to the minima for the complex structure moduli
and the dilaton. Prior to the introduction of the perturbative
corrections to the K\"ahler potential, the complex structure
moduli and the dilaton were fixed by background fluxes through the
conditions $D_UW=0$ and $D_SW=0$. Consider now the case that the
K\"ahler corrections, which are negative near to the minimum of
$T$, try to drive away $S$ and/or $U$ from their minima
$D_SW=D_UW=0$. Then the tree level flux potential yields a
contribution $V_{\rm flux}\sim {\cal O}(+\frac{1}{{\cal V}^2})$,
while the K\"ahler corrections contribute at ${\cal
O}(-\frac{1}{{\cal V}^3})$, which is subleading at large volumes.
Thus, the corrections cannot destabilize the original minima of
$S$ and $U$ and these remain minima of the full theory including
the K\"ahler corrections which stabilize $T$.\footnote{This
argument follows a similar discussion in~\cite{Brama2,CQS} where
the interplay of the ${\cal O}(\alpha'^3)$-correction with the
potential induced by gaugino condensation was studied at large
volume.  Notice that this is an improvement on
KKLT~\cite{nilles2}, where the stabilizing F-term potential from
gaugino condensation is ${\cal O}(-\frac{1}{{\cal V}^2})$.}
Moreover, similar arguments can of course be used after including
the uplifting, to which we now turn.

\section{de Sitter vacua from a consistent D-term}\label{dS}

It is now easy to see that the perturbative $AdS_4$-minimum for
$T$ discussed in the last Section can be uplifted to a
$dS_4$-minimum with a consistent D-term. The full theory including
the perturbative corrections to the K\"ahler potential is a
function of $T+\bar{T}$ alone. Thus it is fully invariant under
the shift $T\to T+i\alpha$ and in particular we have invariance of
$G=K+\ln|W|^2$ under this shift symmetry. Therefore, the mechanism
of K\"ahler stabilization of the volume modulus $T$ fulfills the
consistency constraints of Sect.~\ref{D-term}. This allows us to
gauge the shift symmetry, using world-volume fluxes on a
$D7$-brane, as described in that section.

The full scalar potential will now contain a D-term piece in
addition to the F-term contributions from the K\"ahler
corrections. For the $T^6/\mathbb{Z}_6'$ example the potential,
expanded up to ${\cal O}(\alpha'^3/(T+\bar{T})^{3/2})$ and to
leading order in the string loop corrections, reads \bea
V=V_F+V_D&=&e^{K^{(0)}}|W|^2\cdot\left(\frac{3}{4}\,\xi
\cdot\frac{(S_0+\bar{S}_0)^{3/2}}{(T+\bar{T})^{3/2}}
+\frac{2\beta_2}{\beta}\,
\frac{(U_0+\bar{U}_0)^2}{(T+\bar{T})^2}\right)\nonumber\label{VFD}\\
&\phantom{=}&\;\;+\,\frac{1}{2}\,({\rm Re}\,f_T^{\rm
D7})^{-1}D_T^2\;,\nonumber\\
&\phantom{=}&\hspace{7.ex}f_T^{\rm D7}=T+k\,S\;{\rm
and}\;D_T=\frac{3\,q}{T+\bar{T}}\cdot \left(1+{\cal
O}(\alpha'^3,g_{\rm S}^2)\right)\;\;,\;k\geq0\nonumber\\
&\phantom{=}&\nonumber\\
&=& \frac{|W|^2}{(T+\bar{T})^3}\cdot\left(\frac{3}{4}\,\xi
\cdot\frac{(S_0+\bar{S}_0)^{3/2}}{(T+\bar{T})^{3/2}}
+\frac{2\beta_2}{\beta}\,\frac{(U_0+\bar{U}_0)^2}{(T+\bar{T})^2}\right)\nonumber\\
&\phantom{=}&\;\;+\,\frac{1}{1+k\cdot\frac{S_0+\bar{S}_0}{T+\bar{T}}}
\cdot\frac{9\,q^2}{(T+\bar{T})^3}\eea where $U_0$ and $S_0$ are
constants given by the values of $U$ and $S$ in the minima
determined by $D_UW=D_SW=0$.~\footnote{$k$ is a function of the
gauge flux on the $D7$-brane and vanishes for zero
flux~\cite{Krefl}. We thank D. Krefl for bringing this to our
attention.} Note that $V_D$ has been expanded only to leading
order since later tuning will require $V_D$ to cancel $V_F$ to
leading order. Taking into account the higher orders in $D_T$
would require to write the higher orders in $V_F$ as well for
consistency. The D-term contribution is the one coming from the
$D7$-brane and comparison with its DBI action allows for the
identification \beq q^2\sim
\int_{\Gamma}d^4\xi_y\sqrt{\tilde{g}_8}F_{mn}F^{\widetilde{mn}}
\label{q}\;\;.\eeq Here a tilde denotes the use of the Weyl
rescaled metric showing the dependence on the radial (volume)
modulus $T=e^{4u}+ib$ \beq
ds^2=e^{-6u}\tilde{g}_{\mu\nu}dx^{\mu}dx^{\nu}
+e^{2u}\tilde{g}_{mn}dy^mdy^n\;\;.\label{metric}\eeq

\FIGURE[ht]{\epsfig{file=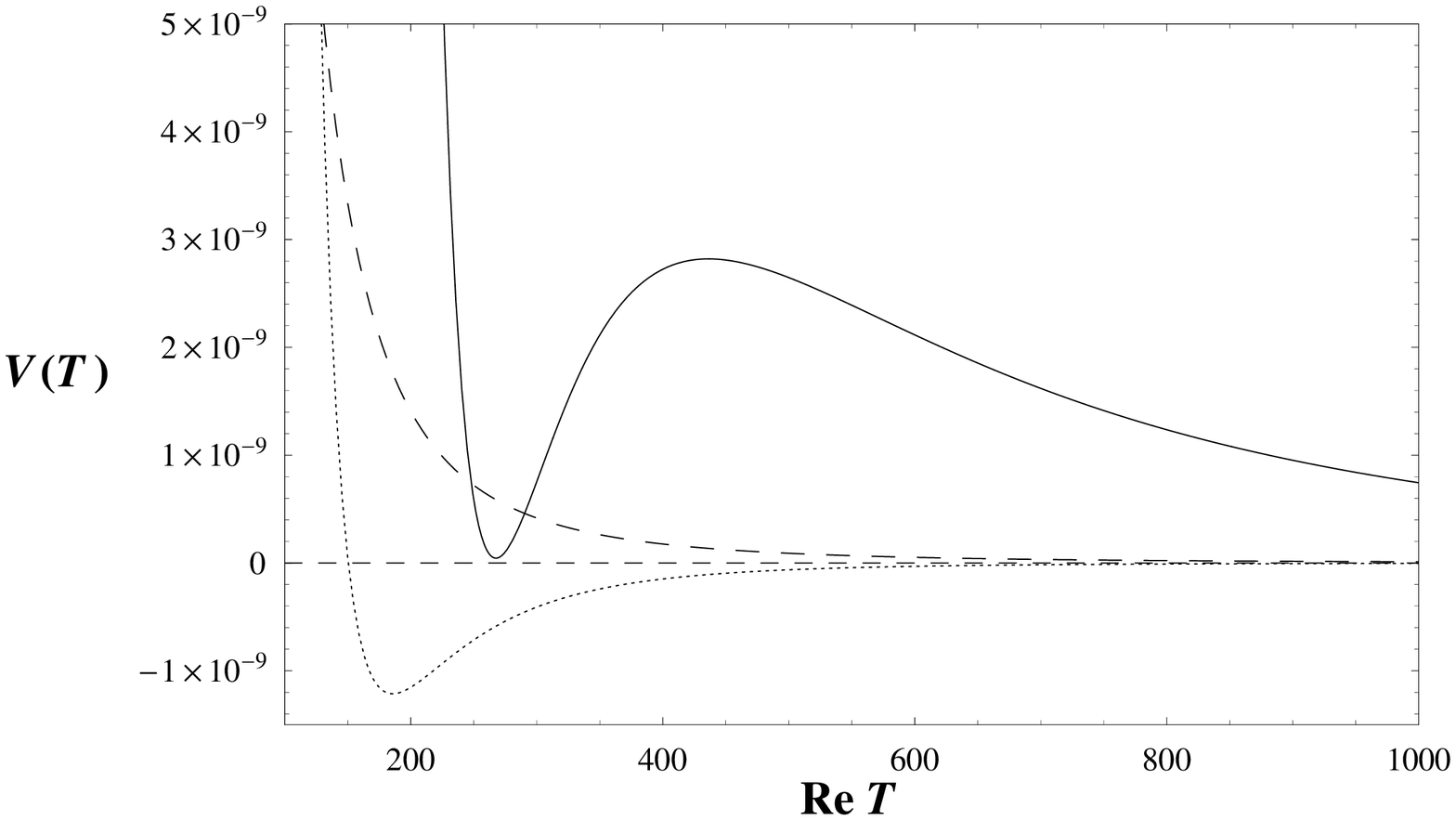,width=12cm} \caption{Dotted: The
F-term scalar potential $V_F(T)$ leading to perturbative K\"ahler
stabilization of $T$. Dashed: The uplifting D-term scalar
potential $V_D(T)$. Both graphs have been rescaled by $10^{-2}$
for display reasons. Solid: The scalar potential eq.~\eqref{VFD}
after uplifting by switching on a gauge field background on a
single $D7$-brane. The numbers are chosen in this example as
$W_0=25.5$, $q=1$, ${\rm Re}\,U=242$, ${\rm Re}\,S=10$, $\chi=48$
and $k=0$ (for simplicity since $V_D$ would get corrected by only
about 5\% for a $k={\cal O}(1)$). Also
$\beta_2=\beta_1=3/256\pi^6$ and $\beta=4\pi^2\beta_2$ are taken
from the $T^6/\mathbb{Z}_2\times\mathbb{Z}_2$ as a guiding
example.}%
\label{Fig.1}}

This identification implies that the values of $q$ are quantized
since the $D7$-brane carries an effective $D3$-charge as seen from
\beq S_{D7}\supset \mu_7\cdot(2\pi\alpha')^2\int_{R^4\times\Gamma}
C_{(4)}\wedge F_{(2)}\wedge F_{(2)} \sim
\underbrace{\mu_7\cdot(2\pi\alpha')^2\cdot\int_{\Gamma}
F_{(2)}\wedge
F_{(2)}}_{\mu_3}\int_{R^4}C_{(4)}\;\;.\label{D3charge}\eeq Since
any change in the $D3$-charge has to be compensated by either
changes in the number of $D3$-branes or discrete flux units this
implies the discreteness of $q$.

Given that ${\rm Re}\,T$ is stabilized at large volume ${\cal V} =
(T + \bar{T})^{3/2}$, and assuming that $q^2$ is ${\cal O}(1)$, we
can arrange for a situation where \beq \Big|\left.V_F\right|_{\rm
min}\Big|\sim \frac{|W|^2}{{\cal V}^3}\sim \left. V_D\right|_{\rm
min}\sim\frac{q^2}{{\cal V}^2}\label{tuning}\eeq holds by tuning
$W$ to larger values such that we get $V_F+V_D\approx 0$ in the
minimum.

This situation is displayed in Fig.~\ref{Fig.1} using the
potential given in eq.~\eqref{VFD} for the semi-explicit
$T^6/\mathbb{Z}_6'$-example. This serves as an indication of how
we expect the behavior to be in a fully explicit model. However,
we should keep in mind that there may be unknown contributions in
this example and the numerical results may change. Moreover, on
the level of the present discussion we have omitted the charged
matter fields of the open string sector which may contribute to
the K\"ahler potential via loops, too, in a complete model. In any
case, given the vast landscape of type IIB flux compactifications,
we expect that there should be many models in type IIB string
theory where our uplifting scenario yields qualitatively the same
results as discussed here.

\section{Conclusion}\label{con}

In this paper we discussed a mechanism for generating de Sitter
vacua in string theory by spontaneously breaking supersymmetry
with consistent D-terms. This proposal has proven difficult to
consistently embed in a stringy scenario.  We find that type IIB
flux compactifications, with volume stabilization via perturbative
corrections to the K\"ahler potential, provide such a scenario. As
discussed in the literature, $\alpha'$- and string loop
corrections allow for stabilization of the $T$ modulus by purely
perturbative means without turning to non-perturbative effects
such as gaugino condensation. Unlike non-perturbative effects, the
K\"ahler corrections preserve the invariance of the theory under a
shift symmetry of the $T$ modulus. In the presence of a magnetised
$D7$-brane this unbroken shift invariance is gauged which leads to
supersymmetric D-terms from string theory which fulfill all the
known consistency requirements of 4d ${\cal N}=1$ supergravity.
These D-terms then provide a parametrically small and tunable
uplift of the perturbatively stabilized $AdS_4$-minimum towards a
metastable $dS$-minimum. In view of the desire to search the
'landscape' of string theory vacua for those regions where
spontaneously broken supersymmetry allows for certain control of
the low-energy effective theory, the discussed mechanism of a
consistent D-term uplift in string theory promises access to a new
class of metastable $dS$-vacua. In a fully explicit model, it
would be necessary to calculate the string loop corrections in the
presence of gauged symmetries and magnetised D-branes, for example
along the lines of \cite{gaugedloops}.  It would also be
interesting to study the consequences of this uplifting mechanism
for possible realizations of inflation in string theory, as well
as the low-energy phenomenology of this type of spontaneous SUSY
breaking.

While this manuscript was being prepared another
paper~\cite{Carlos} appeared which studies the possibility of
consistently realizing the original proposal of~\cite{bkqu} in a
4d KKLT inspired setup.

\acknowledgments

We would like to thank B.~S.~Acharya, M.~Haack, A.~Hebecker,
F.~Marchesano, F.~Quevedo, M.~Serone and I. Zavala for useful
discussions and comments.

\end{document}